\documentclass[a4paper,noarxiv,twocolumn,longbibliography]{quantumarticle}
\usepackage[utf8]{inputenc}
\usepackage[english]{babel}
\usepackage[T1]{fontenc}
\usepackage{lipsum}
\usepackage{mdframed}
\usepackage{booktabs}
\usepackage{longtable}
\usepackage{eurosym}
\usepackage{indentfirst}
\usepackage{amsmath,amssymb,amstext}
\usepackage[colorlinks,bookmarks=false,citecolor=blue,linkcolor=red,urlcolor=blue]{hyperref}
\begin{document}
\title{Time crystal and chaos in the hybrid atom-optomechanics system}
\author{Xingran Xu}
\email{thoexxr@hotmail.com}
\affiliation{Division of Physics and Applied Physics, School of Physical and Mathematical Sciences, Nanyang Technological University, Singapore 637371, Singapore}
\author{Tanjung Krisnanda}
\affiliation{Division of Physics and Applied Physics, School of Physical and Mathematical Sciences, Nanyang Technological University, Singapore 637371, Singapore}

\author{Timothy C.~H. Liew}
\email{TimothyLiew@ntu.edu.sg}
\affiliation{Division of Physics and Applied Physics, School of Physical and Mathematical Sciences, Nanyang Technological University, Singapore 637371, Singapore}
\affiliation{MajuLab, International Joint Research Unit UMI 3654, CNRS, Universit\'e C\^ote d'Azur, Sorbonne Universit\'e, National University of Singapore, Nanyang Technological University, Singapore}

\date{\today}
\begin{abstract}
We consider atoms in two different periodic potentials induced by different lasers, one of which is coupled to a mechanical membrane via radiation pressure force.
The atoms are intrinsically two-level systems that can absorb or emit photons, but the dynamics of their position and momentum are treated classically.
On the other hand, the membrane, the cavity field, and the intrinsic two-level atoms are treated quantum mechanically.
We show that the mean excitation of the three systems can be stable, periodically oscillating, or in a chaotic state depending on the strength of the coupling between them.
We define regular, time crystal, and chaotic phases, and present a phase diagram where the three phases can be achieved by manipulating the field-membrane and field-atom coupling strengths.
The first and second-order correlation functions in different phases are also calculated, which can be observed in experiments.
Our proposal offers a new way to generate and tune time crystal and chaotic phases in a well-established atom-optomechanics system.

\end{abstract}
\maketitle
	
	
\section{Introduction}
The hybrid atom-optomechanics system has been exploited due to its rich physics that allows for many opportunities, from theoretical proposals to experimental implementations.
Frequent configuration of the system consists of a mechanical membrane (oscillator) and a Bose-Einstein condensate (BEC) that are mutually coupled to cavity field modes~\cite{Jockel2015,Mann20182,Brennecke_2008,Pirkkalainen_2015}.
Applications resulting from this system have been valuable.
For example, the mechanical oscillator can be cooled down by enhancing the effective coupling strength between the membrane and the atom~\cite{Vogell_2015,Mann20182,Jockel2015,Bariani2014,Vogell2013,Vochezer2018}.
At the same time, the BECs can have a nonequilibrium phase transition from the normal phase to the self-organized super-radiant phase~\cite{Mann2018,Bakhtiari2015,Xu_2019,Klinder2015,Gao2019} due to the $Z_2$ symmetry breaking~\cite{Cheng_2021}.
The system is also applicable for metrology~\cite{Lau_2019,Cronin2009,Aspelmeyer2014} and quantum simulations~\cite{Bloch_2012,Manukhova_2020}.
Last but not least, it provides a new platform to create new states of many-body physics, such as the spontaneous crystallization of atoms and light into a structure that features phonon-like excitations and bears similarities to a supersolid~\cite{Ostermann2016,Schuster2021,Baio2021,Mivehvar2018,Nagy2008}.
This motivates the further study of this system to potentially realize marvelous dynamical phases such as time crystal and chaos.

The time crystal phase breaks the time-translation symmetry~\cite{Timec1,Timec2}, which is beyond the strict thermal equilibrium~\cite{Else2016,Bruno2013,Bruno20132,Watanabe2015}.
Indeed, while quantum time crystals were originally defined as systems whose lowest energy state undergoes periodic motion~\cite{Wilczek2012}, the definition has been extended to include nonlinear driven-dissipative systems~\cite{Colas2016,Nalitov2019}.
Time crystals have been observed in many nonequilibrium experiments such as driven disordered dipolar spin impurities in diamond~\cite{Choi_2017}, the interacting spin chain of trapped atomic ions~\cite{Zhang_2017}, quantum computing processor~\cite{mi2021observation}, etc. On the other hand, the chaos phase represents unpredictable results after a long evolution time, which are sensitive to initial states~\cite{Eckmann1985}.
In this case, chaotic attractors may arise, leading to orbits that converge to the corresponding chaotic region in the phase-space diagram~\cite{Lorenz_1963,Strelioff2006}.
For optomechanical system, chaotic dynamics appears in the bad-cavity limit and is described by the semiclassical equations of motion~\cite{Bakemeier2015,Yang_2019,Montoya_2019}.
The hybrid atom-optomechanics system can be conditioned such that it satisfies the requirements for both time crystal and chaotic phases, which urges for a proposal for their realization.

Experimental implementations have been reported for the atom-optomechanics system with $^{87}$Rb atoms and Si$_3$N$_4$ or SiN membrane~\cite{Camerer2011,Jockel2015}, where the position of the membrane displaces the lattice potential for the atoms~\cite{Bloch_2005,Hammerer2010,Christoph2018,Bennett2014}.
Meanwhile, the center-of-mass motion of the atoms will experience a restoring optical dipole force due to the absorption and stimulated emission~\cite{Camerer2011,Mann2018,Bakhtiari2015,Asboth2008}.
The optical lattice for the atoms can be highly engineered with different potentials.
The depth of the potential can be adjusted by the power of the laser, while the period can be tuned by changing the wavelength of the laser or the angle between two beams~\cite{Fallani_2005,Bloch_2012,Bloch_2005}.
The effective coupling between the atoms and the membrane can be long distance interaction mediated by the laser field.
The field interacts with the atoms via light-matter coupling, with an effective strength enhanced by the number of atoms ($\sim 10^{10}$)~\cite{Jockel2015,Vogell2013}.

In this paper, we consider the atoms trapped in a double well-like potential created by two lasers, where the wavelength of one is half of the other.
One of the lasers is filtered and enters a cavity where it couples to a membrane, which in turn affects its optical path.
This way, the atoms will have both time-dependent and fixed potentials.
In this configuration, relevant interactions include optomechanical coupling between the field and the membrane as well as light-matter coupling between the field and the atoms.
The position and momentum of the atoms are treated classically, with their dynamical equations coupled to a quantum master equation characterizing the membrane, the cavity field, and the intrinsic degrees of freedom of the atoms (two-level systems).
We show that by tuning the strength of the optomechanical and light-matter coupling, the system can be in regular, time crystal, or chaotic phases.
We also computed experimentally familiar quantities such as the first and second-order correlation functions in different phases.
\section{Model}
 
\begin{figure}[h!]
\centering
\includegraphics[width=0.5\textwidth]{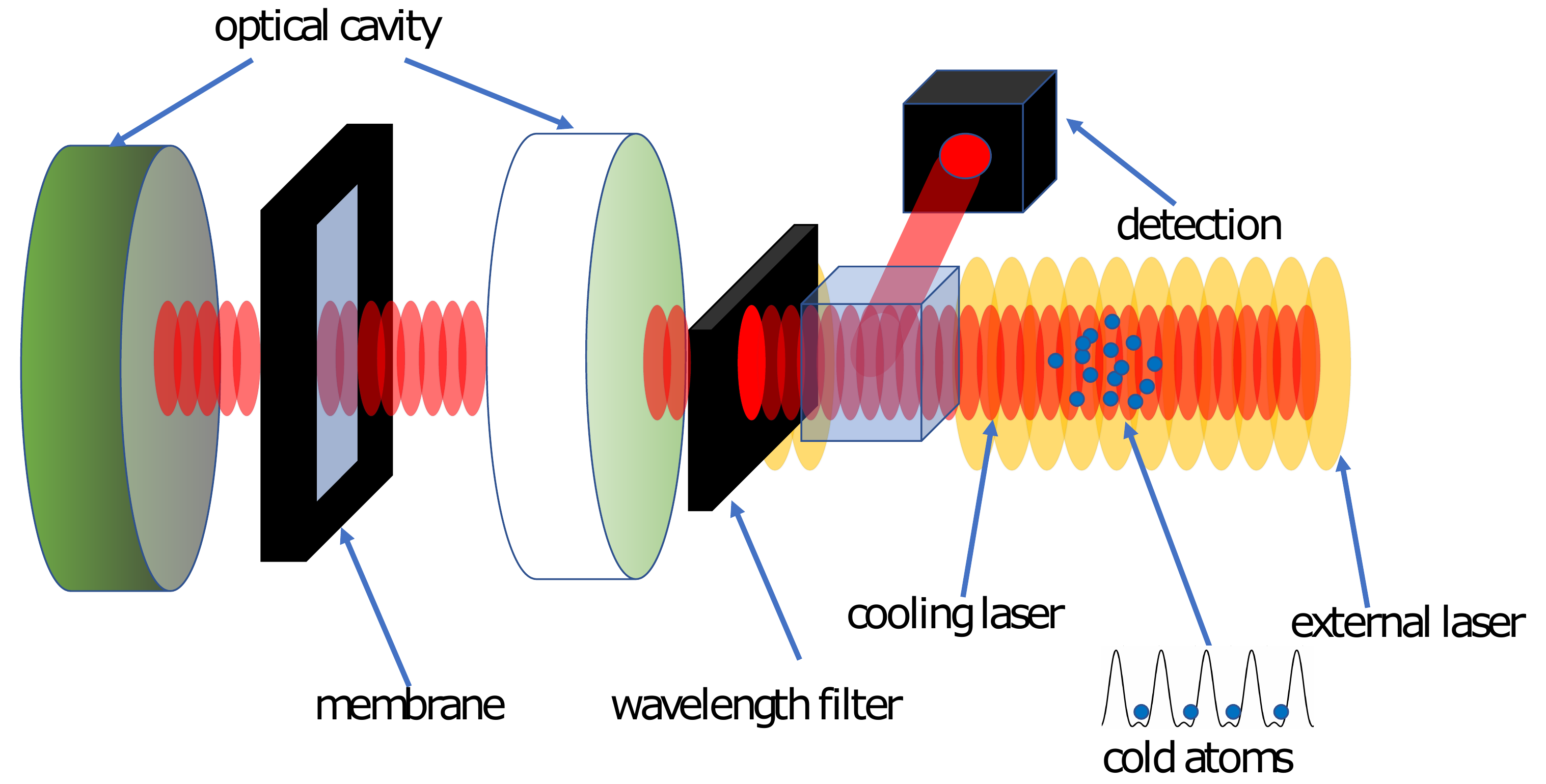}\\
\caption{The scheme of the hybrid atom-optomechanics system. The cold atoms are trapped by two lasers and only one of them is filtered to enter a cavity where it is coupled to a mechanical membrane.}\label{shiyi}
\end{figure}

Consider two-level atoms moving in an adjusted gauge field optical lattice, which is coupled to a membrane through the coherently driven cavity field as shown in Fig.~\ref{shiyi}.
The cold atoms are trapped by two lasers with different wavelengths, giving two optical lattice potentials with different periods.
The Hamiltonian ($\hbar=1$) describing the membrane, cavity field, and two-level atom, in a frame rotating with the driving frequency $\omega_l$ and with rotating-wave approximation, is written as
\begin{eqnarray}
H &=& \omega_m \hat b^{\dagger}\hat b-\Delta_c \hat a^\dagger \hat a - \Delta_a\hat \sigma^+ \hat \sigma^- +\eta(\hat a + \hat a^\dagger)\nonumber \\
&-&g_{mc}(\hat b^\dagger+\hat b)\hat a^{\dagger}\hat a + g_{ac}\sin(2x)(\hat a^\dagger \hat \sigma^- + \hat a\hat \sigma^+),
\end{eqnarray}
where $\omega_m$ is the frequency of the membrane, $\Delta_c = \omega_l - \omega_c$ and $\Delta_a=\omega_l - \omega_a$ are the detuning for the cavity and atom, respectively.
$\omega_c$ denotes the cavity frequency, $\omega_a$ the atomic transition frequency and $\omega_l$ the frequency of the laser driving the cavity with strength $\eta$.
The atom couples to the cavity (Jaynes-Cummings type) with strength $g_{ac}$, while the optomechanical coupling~\cite{Aspelmeyer2014} between the membrane and cavity is denoted by $g_{mc}$.
The optical lattice has a mode function $\sin(2x)$, where $x$ is the atomic position, which is in units of the inverse cavity wave number.
The annihilation operators for the cavity, atom, and membrane are denoted by $\hat a$, $\hat \sigma^-$, and $\hat b$, respectively.

\begin{figure*}
\centering
\includegraphics[width=0.85\textwidth]{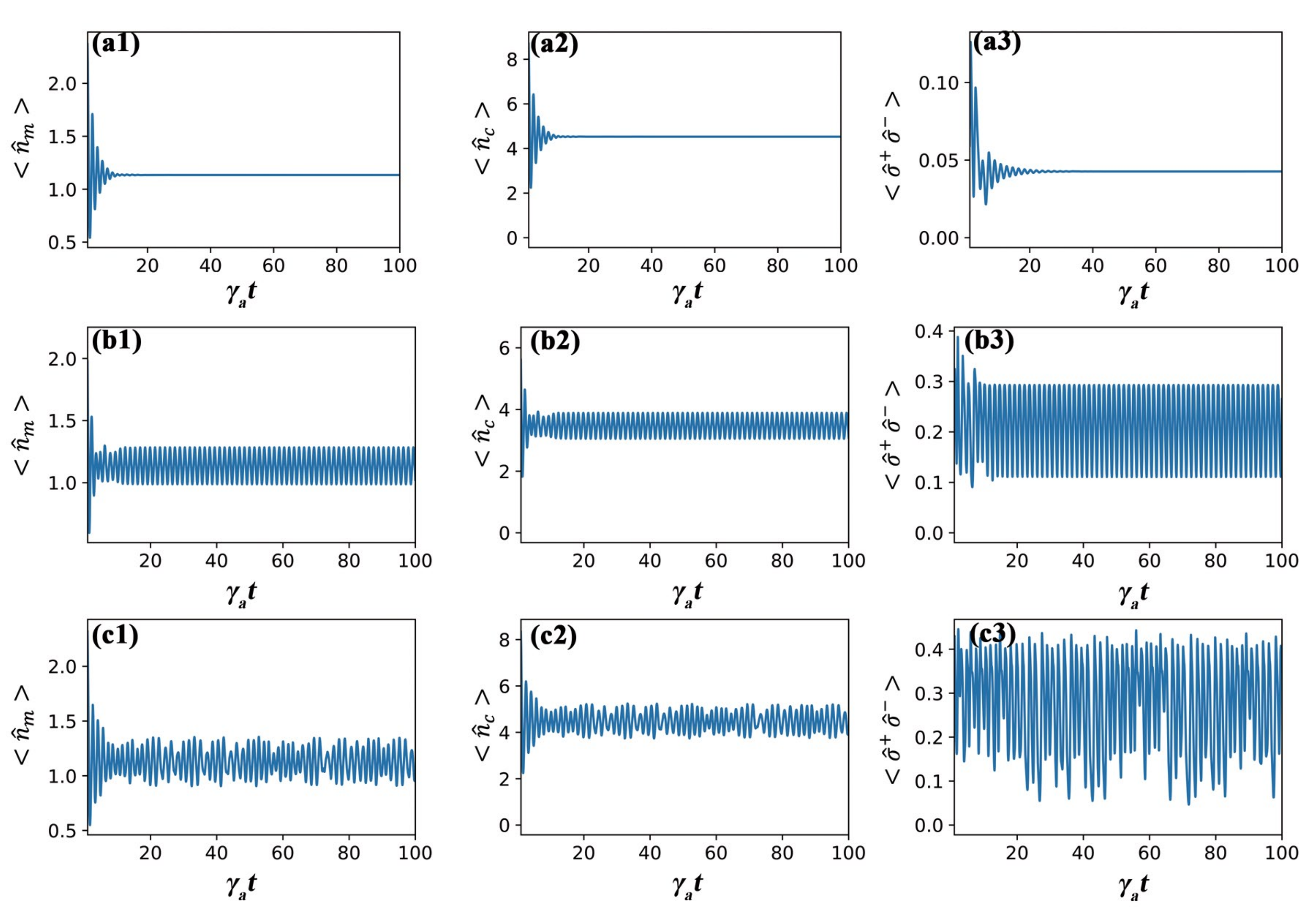}\\
\caption{ The mean excitation of the membrane mode (first column), the cavity mode (second column) and the atom (third column).
The parameters used are $\gamma_c/\gamma_a$=0.5, $\gamma_m/\gamma_a$=2, $\eta/\gamma_a$=5, $V_0/\gamma_a$=20, $V_1/\gamma_a$=40, $\omega_r/\gamma_a$=1, $\Delta_c/\gamma_a$=-1, and $\Delta_a/\gamma_a$=-2.
Specific coupling parameters for different phases are given by (a1)-(a3): $g_{ac}/\gamma_a$=0.5, $g_{mc}/\gamma_a$=2; (b1)-(b3): $g_{ac}/\gamma_a$=2, $g_{mc}/\gamma_a$=2.5; and (c1)-(c3): $g_{ac}/\gamma_a$=2, $g_{mc}/\gamma_a$=2.}\label{osc}
\end{figure*}

The decays in the system are modelled by the Liouvillians and can be considered as Lindblad terms
\begin{equation}
\mathcal{L}_{\mu}[\rho,\hat O] = \gamma_{\mu}(2\hat O\rho \hat O^\dagger - \hat O^\dagger \hat O \rho - \rho \hat O^\dagger \hat O),
\end{equation}
where $\gamma_{\mu}$ is the dissipation rate of the membrane ($\mu=m$), cavity field ($c$), and two-level atom ($a$).
Note that $\hat O$ denotes the corresponding annihilation operator of each system.
As the initial state, we use uncorrelated states of the form $\rho=\rho_m\otimes\rho_c\otimes\rho_a$, where $\rho_m$, $\rho_c$, and $\rho_a$ represent the density matrix for the membrane, cavity field, and atom.
The evolution follows the quantum master equation:
\begin{equation}\label{EQ_QME}
\dot \rho=-i[H,\rho]+\frac{1}{2}\left(\mathcal{L}_{m}[\rho,\hat b]+\mathcal{L}_{c}[\rho,\hat a]+\mathcal{L}_{a}[\rho,\hat \sigma^-]\right).
\end{equation}
In addition, we have classical differential equations of the atomic motion obtained from the Ehrenfest theorem: $\dot{x}={\partial \langle H\rangle}/{\partial p}$ and $\dot{p}=-{\partial \langle H\rangle}/{\partial x}$, where the observables $x$ and $p$ are treated simply as numbers in the classical regime.
For this classical motion, the atom is situated in two potentials such that its Hamiltonian reads
\begin{equation}\label{EQ_H_atom}
H_a=\frac{p^2}{2m}+V_0\sin(x)^2+V_1 \sin(x),
\end{equation}
where $m$ is the mass of the atom, $V_0$ is the depth of the optical lattice, and $V_1 \sin(x)$ is the external periodic potential.
We stress that only one of the potentials ($V_0$) ends up being coupled to the membrane.
The equations of motion for the atom, taking into account the Hamiltonians $H$ and $H_a$, are written as
\begin{eqnarray}
\dot{x} &=& 2\omega_r p, \nonumber \\
\dot{p} &=& -4g_{ac}\cos(2x)~\Re\left\{\langle \hat a^\dagger \hat \sigma^-\rangle\right\} \nonumber \\
&&-\left[V_1 \cos(x)+V_0\sin(2x)\right],
\end{eqnarray}
where $\omega_r=1/(2m)$ is the recoil frequency.
As initial conditions, we take $x(0)=-1$ and $p(0)=0$. We note that the quantum observable $\Re\left\{\langle \hat a^\dagger \hat \sigma^-\rangle\right\}$ updates the classical dynamics, while the latter affects the quantum dynamics via the change in the optical path, and hence, the mode function $\sin(2x)$.
We show below with suitable parameters, that this quantum-classical coupled dynamics can produce regular, time crystal, and chaotic phases.
See also the Appendix for calculations of the dynamics using the quantum trajectory method.


\section{Different dynamical behaviors}

\begin{figure}[h!]
\centering
\includegraphics[width=0.47\textwidth]{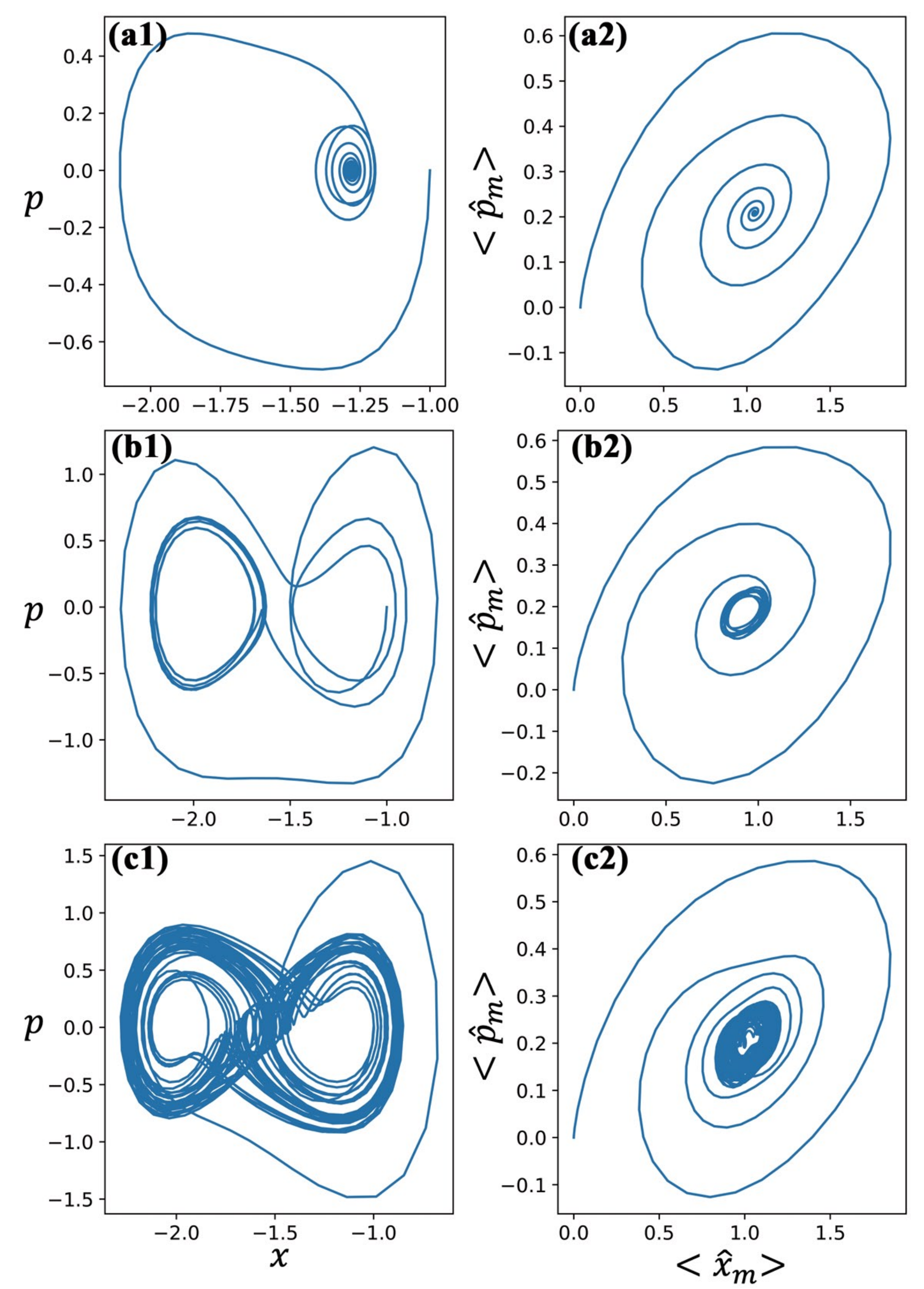}\\
\caption{ The motion of the atom (first column) and the membrane (second column) in the phase-space diagram.
The parameters are given by $\gamma_c/\gamma_a$=0.5, $\gamma_m/\gamma_a$=2, $\eta/\gamma_a$=5, $V_0/\gamma_a$=20, $V_1/\gamma_a$=40, $\omega_r/\gamma_a$=1, $\Delta_c/\gamma_a$=-1, and $\Delta_a/\gamma_a$=-2.
For panels (a1)-(a2): $g_{ac}/\gamma_a$=0.5, $g_{mc}/\gamma_a$=2; (b1)-(b2): $g_{ac}/\gamma_a$=2, $g_{mc}/\gamma_a$=2.5; and (c1)-(c2): $g_{ac}/\gamma_a$=2, $g_{mc}/\gamma_a$=2.}\label{phasemotion}
\end{figure}

The hybrid system has quantum and classical parts that are treated differently but are coupled to each other.
For the quantum part, all three systems are also coupled, and consequently, we note that a particular phase in one system is an indication of the same phase in others.
In what follows, we define three phases based on the dynamical behavior of mean excitations (either of the membrane $\langle n_{m}\rangle$, cavity field $\langle n_{c}\rangle$, or two-level atom $\langle \hat \sigma^+ \hat \sigma^- \rangle$):
\begin{enumerate}
\item \emph{Regular phase.}~The excitation of each system will have a stable value after a long evolution time, as shown in Figs.~\ref{osc}(a1)-(a3).
In this case, the motion of the atom and the membrane will converge to a point in the phase-space diagram, as shown in Figs.~\ref{phasemotion}(a1)-(a2).

\item \emph{Time crystal phase.}~The excitation of each system will oscillate periodically around a certain value, as shown in Figs.~\ref{osc}(b1)-(b3).
Here, after a certain time, the motion of the atom and the membrane will continue to orbit a point in the phase-space diagram, see Figs.~\ref{phasemotion}(b1)-(b2).

\item \emph{Chaotic phase.}~The excitation of each system will show random oscillation, as shown in Figs.~\ref{osc}(c1)-(c3).
The motion of the atom and the membrane in the phase-space diagram will exhibit random orbits around two attractors, see Figs.~\ref{phasemotion}(c1)-(c2).
\end{enumerate}

Remarkably, the quantum-classical coupled dynamics can produce all three phases, by simply tuning the strength of the optomechanical ($g_{mc}$) and light-matter ($g_{ac}$) coupling, see Fig.~\ref{osc}.

Furthermore, the atomic motion ($x,p$) and expectation value of quadratures, e.g., for the membrane $(x_m, p_m)$, where $x_{m}=\langle \hat b+\hat b^{\dagger}\rangle/\sqrt{2}$ and $p_{m}=\langle \hat b-\hat b^{\dagger}\rangle/(i\sqrt{2})$ are plotted in phase-space diagrams in Fig.~\ref{phasemotion}.
For the atom, as the period of one potential is twice the other (see Eq.~(\ref{EQ_H_atom})), the depths $V_0$ and $V_1$ allow for a double well-like potential shape, which consequently gives three optimum points, two of which are stable.
The position of the two stable points are symmetric with respect to, e.g., $x=-\pi/2$ where the stronger potential $V_1\sin(x)$ has the lowest energy.
Thus, the steady momentum is always zero in the regular phase, whereas it is oscillating around zero in other phases.
The atomic motion will converge to one of the stable points in the regular phase while the trajectory will form a closed circle in the time crystal phase.
When the system is in the chaotic phase, there are two attractors in the phase-space diagram and the motion of the atom is unpredictable.
At the same time, the behaviors of the quantum degrees of freedom reflect that of the classical ones ($x,p$) of the atom, see the second column of Fig.~\ref{phasemotion} for the membrane's quadratures.
The motion of the membrane will have non-zero momentum in the steady-state regime.
Below we shall introduce quantities to indicate the phase of the system, and finally, obtain a phase transition diagram for varying values of the optomechanical and light-matter coupling strengths.

We also computed the first and second order correlation functions $G^{(1)}(\tau)$ and $G^{(2)}(\tau)$ that are standardly measured in experiments.
See the Appendix for details.
As expected, the behaviors of these correlation functions follow that of the mean excitation in the corresponding phases.

\begin{figure}[h!]
\centering
\includegraphics[width=0.5\textwidth]{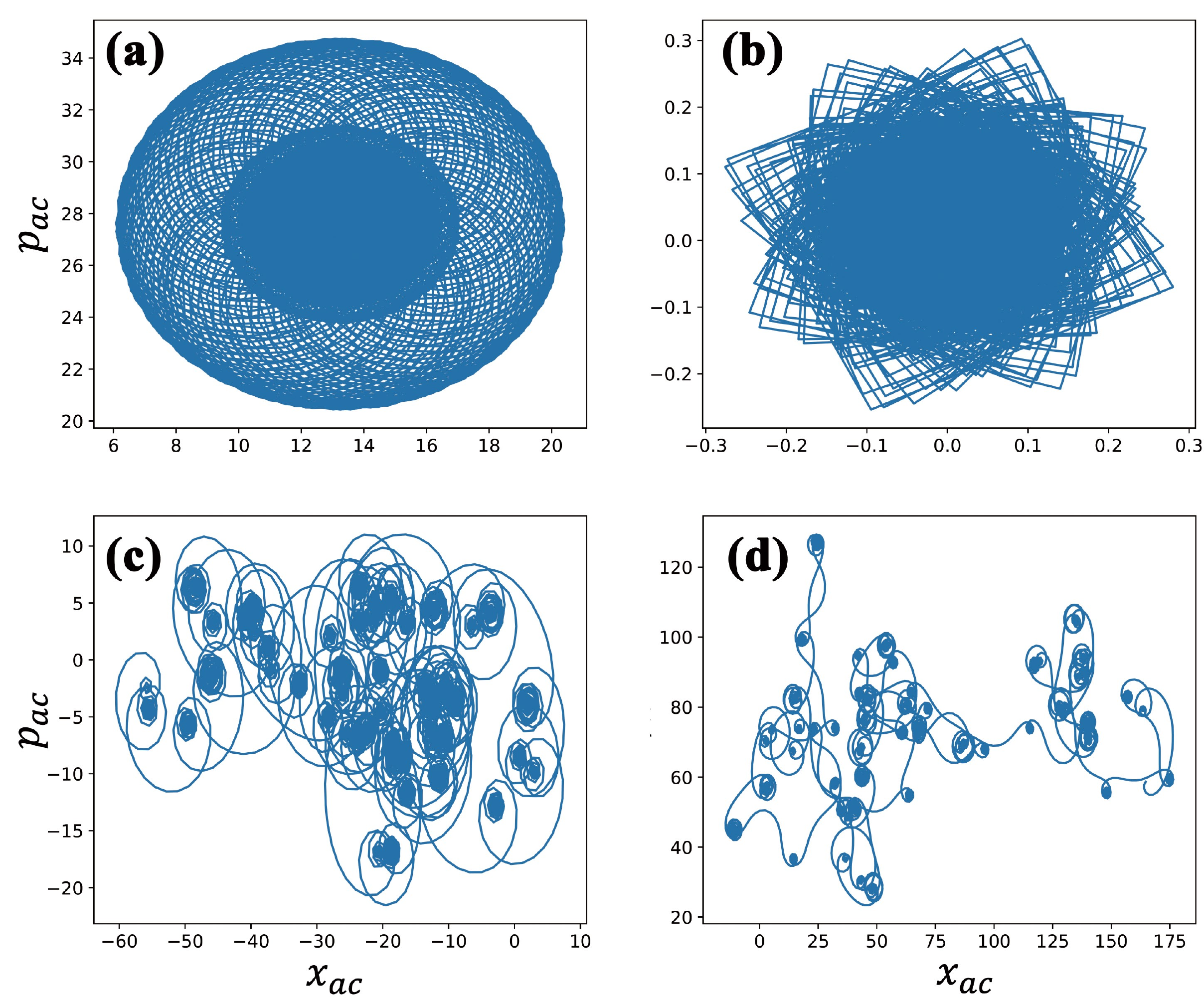}\\
\caption{The dynamics of the translation components $(x_{ac},\:p_{ac})$.
The parameters used are $\gamma_c/\gamma_a$=0.5, $\gamma_m/\gamma_a$=2, $\eta/\gamma_a$=5, $V_0/\gamma_a$=20, $V_1/\gamma_a$=40, $\omega_r/\gamma_a$=1, $\Delta_c/\gamma_a$=-1, and $\Delta_a/\gamma_a$=-2. We also used (a): $g_{ac}/\gamma_a$=0.5, $g_{mc}/\gamma_a$=2; (b): $g_{ac}/\gamma_a$=2, $g_{mc}/\gamma_a$=2.5; (c): $g_{ac}/\gamma_a$=2, $g_{mc}/\gamma_a$=2; and (d): $g_{ac}/\gamma_a$=4, $g_{mc}/\gamma_a$=2}\label{fractal}
\end{figure}

\section{Quantification and classification of the phases}

Here we shall present a way to numerically classify the phases previously described.
In particular, we used two quantities, where one is recognizing the regular phase and the other the chaotic phase.
Consequently, this method classifies all three possible phases in the phase transition diagram, which we will present below.

\emph{The regular phase transition}.
Recognizing the regular phase is straightforward as the mean excitation of all the systems will go towards a constant value, see Fig.~\ref{osc}.
Here, after a long evolution time, one can choose a time range and compute $R=\max{(\langle n_{\mu}\rangle)}-\min{(\langle n_{\mu}\rangle)}$.
The regular phase is given for $R<\epsilon$, where $\epsilon$ is a small constant.

\emph{The 0-1 test for the chaotic phase transition}.
The system in the chaotic phase will have a very different dynamical behavior, which can be tested by the regression or correlation method~\cite{Gottwald_2009}.
Here, relevant functions are defined such that we can apply the above tests to our system.
First, we take new translation components ($x_{ac}$,\:$p_{ac})$ and $\theta_c$ as follows
\begin{eqnarray}
p_{ac}(n+1)&=&\phi(n) \cos(\theta_c)+p_{ac}(n),\nonumber \\
x_{ac}(n+1)&=&\phi(n)\sin(\theta_c)+x_{ac}(n+1),\nonumber \\
\theta_c(n+1)&=&\nu+\theta_c(n)+\phi(n),
\end{eqnarray}
where $n=1,2,\cdots, N$ denotes the time index, $\phi(n)$ is a dynamical quantity, here taken as $x(n)+p(n)$ , and $\nu$ is a fixed constant $[0,\pi]$. The initial state of $p_{ac}$, $x_{ac}$ and $\theta_c$ are zero and they are updated by the position and the momentum of atoms.
The quantities $q_{ac}$ and $p_{ac}$ are bounded if the dynamical behavior is regular, while in the chaotic phase they will behave asymptotically.
The translation components resulting from the hybrid atom-optomechanical system are shown in Fig.~\ref{fractal}.
The regular and time crystal phases have bounded states for ($x_{ac},p_{ac}$) as shown in Figs.~\ref{fractal}(a) and (b).
However, they become unbounded in the chaotic phase, see Figs.~\ref{fractal}(c) and (d), showing the pattern of fractals.

Given dynamical components ($x_{ac},p_{ac}$), the mean square displacement is defined as
\begin{eqnarray}
M_c(n)&=&\lim_{N\rightarrow\infty}\frac{1}{N}\sum_{j=1}^{N}\left[p_{ac}(j+n)-p_{ac}(j) \right]^2 \nonumber \\
&&+\left[ x_{ac}(j+n)-x_{ac}(j) \right]^2,
\end{eqnarray}
where $n\ll N$ is required.
The test for chaos is based on the growth rate of $M_c(n)$ as a function of $n$.
A modified mean square displacement $D_c(n)$ that exhibits the same asymptotic growth as $M_c(n)$, but with better convergence properties is given by
\begin{equation}
D_c(n)=M_c(n)-V_{osc}(\nu,n),
\end{equation}
where the oscillation term $V_{osc}$ is defined as $V_{osc}=(E_\phi)^2({1-\cos(n\nu)})/({1-\cos(\nu)}),$
and the expectation $E_\phi$ is given by $E_\phi=\lim_{N\rightarrow\infty}\frac{1}{N}\sum_{j=1}^N \phi(j)$.
Note that the cut-off index $n_{\text{cut}}$ needs to be large enough such that the error of $D_c(n_{\text{cut}})$ is close to zero.

\begin{figure}[h!]
\centering
\includegraphics[width=0.5\textwidth]{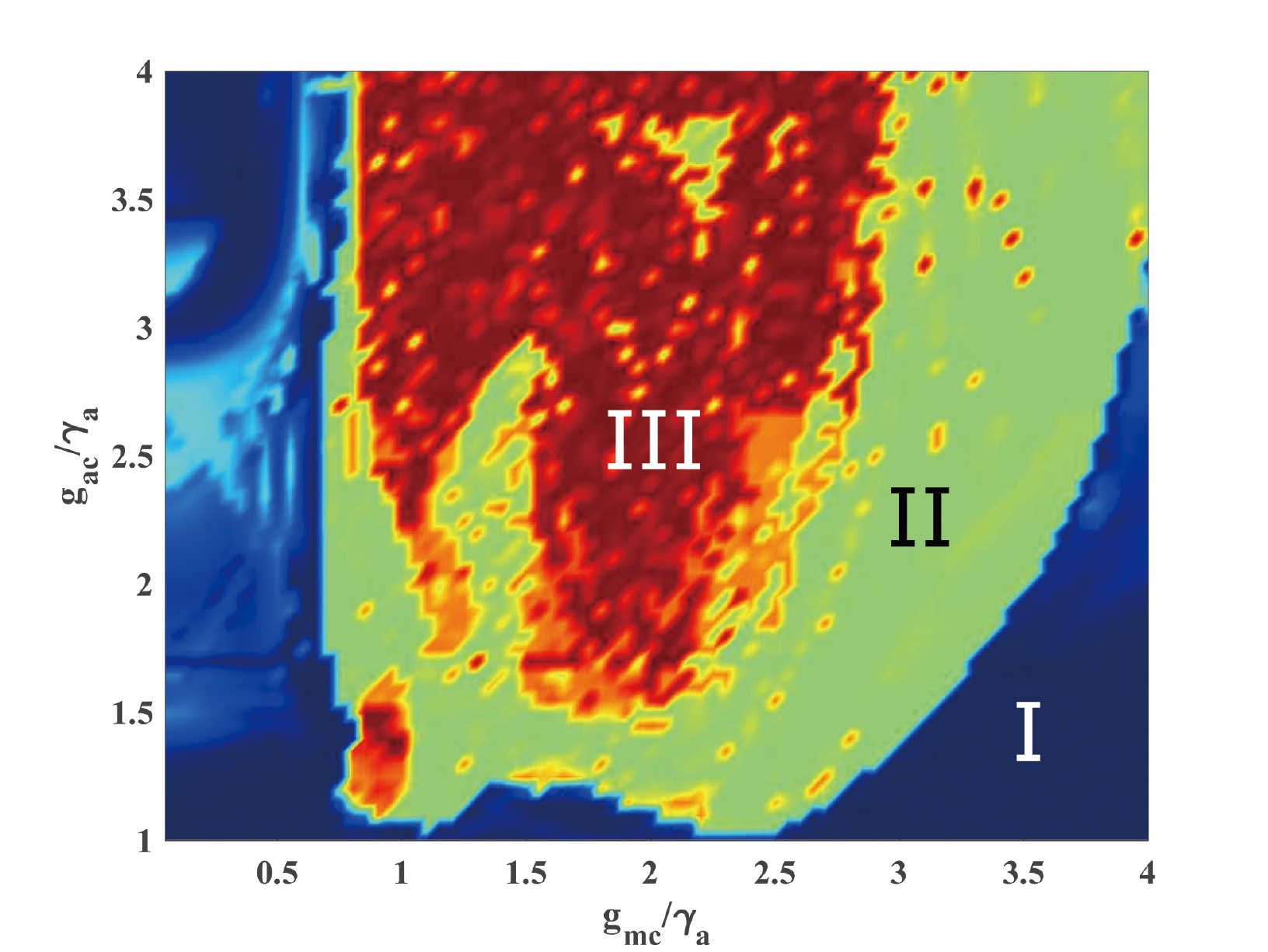}
\caption{The phase transition diagram. 
The regular, time crystal and chaotic phases are denoted by I, II and III, respectively.
The parameters used are $\gamma_c/\gamma_a$=0.5, $\gamma_m/\gamma_a$=2, $\eta/\gamma_a$=5, $V_0/\gamma_a$=20, $V_1/\gamma_a$=40, $\omega_r/\gamma_a$=1, $\Delta_c/\gamma_a$=-1, and $\Delta_a/\gamma_a$=-2. }\label{phad}
\end{figure}

The 0-1 test via regression method is calculated following the quantity $K_c=\lim_{n\rightarrow \infty}{\log M_c(n) }/{\log n}$,
whose value is near zero (one) for non-chaotic (chaotic) phase.
An alternative test, that we also consider, is via the correlation method~\cite{Gottwald_2009} and it is determined by the mean square displacement $D_c$ as follows
\begin{eqnarray}
\text{cov}(X,Y)&\equiv&\frac{1}{q}\sum_{j=1}^{q}\left( X(j)-\bar{X} \: \right) \left( Y(j)-\bar{Y} \right), \nonumber \\
K_c&=&\frac{\text{cov}(\xi,\Delta)}{\sqrt{\text{cov}(\xi,\xi)\text{cov}(\Delta,\Delta)}},  \label{corm}
\end{eqnarray}
where $\bar{X}$ and $\bar{Y}$ are the mean values of the vectors $X$ and $Y$ with length $q$.
We take the vectors $\xi =(1,2,\cdots,n_{\text{cut}})$ and $\Delta =(D_c(1), D_c(2),\cdots,D_c(n_{\text{cut}}))$.

\emph{The phase transition diagram}.
The three phases are characterized by the two tests described above (regular and chaotic phase transition tests).
The phase diagram for the two tests are plotted separately, see the Appendix.
Here, we combine the diagrams, see Fig.~\ref{phad}, which shows the three phases for different coupling strengths.
When the atom-cavity coupling is close to zero, only the regular phase exists with the balance of the rates of the decay and the drive. 
With the increase of $g_{ac}$, the time crystal phase will appear with periodic evolution of the interaction strength $g_{ac}\Re\left\{\langle \hat a^\dagger \hat \sigma^-\rangle\right\}$. 
For further increase of $g_{ac}$ the system reaches the chaotic phase. 
Remarkably, the coupling between the cavity and the membrane $g_{mc}$ also plays an important role in the time crystal and chaotic phases. 
If $g_{mc}$ is too small compared to $g_{ac}$ the model can be simplified to an atom cooling model and the membrane's oscillations can be ignored. 
On the contrary, if $g_{ac}$ is too small, the system can be transformed to an optomechanical model and the atoms can be ignored. 
The competition of the coupling strengths allows the system to have a rich phase diagram.

\section{Conclusion}
We theoretically considered a hybrid atom-optomechanics system to realize different dynamical phases by exploiting the competition of the coupling strength of the cavity and atoms, and that of the cavity and membrane. 
The atoms experience two potentials, including one that may be static, periodically oscillating, or randomly oscillating. 
The coupling of the cavity mode and the membrane allows them to have similar behavior, where the whole system can exhibit a regular, time crystal, or chaotic phase. 
These three phases are distinguished after evolving quantities from the system for a sufficiently long time, where we performed regular and chaotic phase transition tests.

\section{Acknowledgements}
This work was supported by the Singaporean Ministry of Education, via the Tier 2 Academic Research Fund project MOE2019-T2-1-004.

\section{Appendix}

\subsection{Quantum trajectory method}
We also use the quantum trajectory (QT) method as a separate way to evolve the atom-optomechanics system.
For a review on quantum trajectories, see Refs.~\cite{carmichael2009open,daley2014quantum}.
As described in the main text, the evolution of the system is governed by the coupled quantum-classical dynamics.
The quantum dynamics is described within the quantum master equation, which here we describe using the QT method. 
The observable $\Re\left\{\langle \hat a^\dagger \hat \sigma^-\rangle\right\}$ obtained from the QT will then update the classical dynamics for the atomic motion, which in turn affects the Hamiltonian (via $\sin(2x)$) of all the trajectories.

We begin by noting that the quantum master equation in Eq.~(\ref{EQ_QME}) can be rewritten as 
\begin{equation}\label{EQ_QTEQ}
\dot \rho=-i(H_{\text{eff}}\rho-\rho H_{\text{eff}}^\dagger)+\tilde b \rho \tilde b^\dagger+\tilde a \rho \tilde a^\dagger+\tilde \sigma^- \rho \tilde \sigma^+,
\end{equation}
where $H_{\text{eff}}=H-(i/2)(\tilde b^\dagger \tilde b+\tilde a^\dagger \tilde a+\tilde \sigma^+ \tilde \sigma^-)$ and the decay rates are absorbed into the operators, i.e., $\tilde b=\sqrt{\gamma_m}\hat b$, $\tilde a=\sqrt{\gamma_c}\hat b$, and $\tilde \sigma^-=\sqrt{\gamma_a}\hat \sigma^-$.
The interpretation of Eq.~(\ref{EQ_QTEQ}) is that the system is evolved under $H_{\text{eff}}$ and at the same time possible \emph{jumps} may occur, from the rest of the terms.
This way, the evolution of each trajectory from $t$ to $t+\delta t$ is constructed as follows.
A candidate state is calculated as $|\psi^{(1)}(t+\delta t)\rangle=(\mathbbm{1}-iH_{\text{eff}}\delta t)|\psi(t)\rangle$. 
As $H_{\text{eff}}$ is not Hermitian, one obtains 
\begin{equation}
    \langle \psi^{(1)}(t+\delta t)|\psi^{(1)}(t+\delta t)\rangle=1-\delta p,
\end{equation}
where $\delta p$ is a probability.
One can further note that 
\begin{eqnarray}
    \delta p&=&\delta t\langle \psi(t)|i(H_{\text{eff}}-H_{\text{eff}}^\dagger)|\psi(t)\rangle \nonumber \\
    &=&\delta t \langle \psi(t)|\tilde b^\dagger \tilde b+\tilde a^\dagger \tilde a+\tilde \sigma^+ \tilde \sigma^-|\psi(t)\rangle \nonumber \\
    &=& \delta p_m+\delta p_c+\delta p_a,
\end{eqnarray}
where we have used, e.g., $\delta p_m\equiv \delta t \langle \psi(t)|\tilde b^\dagger \tilde b |\psi(t)\rangle$.
The stochastic evolution step is computed as follows:
\begin{enumerate}
    \item With probability $1-\delta p$, the new state is
    \begin{equation}
        |\psi(t+\delta t)\rangle=\frac{|\psi(t+\delta t)\rangle}{\sqrt{1-\delta p}}.
    \end{equation}
    \item With probability $\delta p$, one of the jumps happens. The new state will be one of the following:
    \begin{eqnarray}
        |\psi(t+\delta t)\rangle&=&\frac{\tilde b|\psi(t)\rangle}{\sqrt{\delta p_m/\delta t}};\nonumber \\
        |\psi(t+\delta t)\rangle&=&\frac{\tilde a|\psi(t)\rangle}{\sqrt{\delta p_c/\delta t}}; \nonumber \\
        |\psi(t+\delta t)\rangle&=&\frac{\tilde \sigma^-|\psi(t)\rangle}{\sqrt{\delta p_a/\delta t}}.
    \end{eqnarray}
    The probability of each state is proportional to $\delta p_m$, $\delta p_c$, and $\delta p_a$, respectively.
\end{enumerate}
The expectation value of an observable is obtained from the average of all trajectories, e.g., 
\begin{equation}
    \langle n_c(t) \rangle=\frac{1}{N}\sum_j^N \langle \psi_j(t)|\hat a^\dagger \hat a|\psi_j(t)\rangle,
\end{equation}
where $|\psi_j(t)\rangle$ is the state of the $j$th trajectory.
For initial states that are mixed, pure states are sampled from the ones composing the initial density matrix, which are then evolved following the QT method.

\begin{figure}[!h]
\centering
\includegraphics[width=0.45\textwidth]{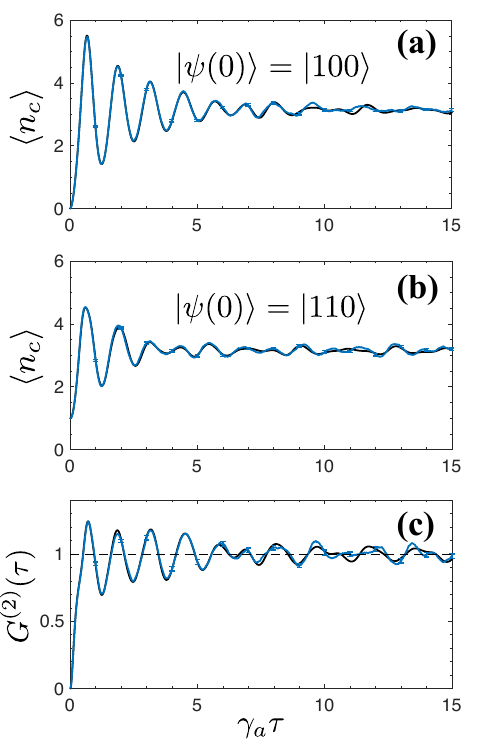}\\
\caption{The evolution of mean excitation of the cavity field mode via the quantum trajectory method. 
Panels (a) and (b) represent evolution starting with different initial states.
Panel (c) is the second order correlation function.
The corresponding results using the quantum master equation are also plotted in each panel (solid black curves).
The error bars represents the standard error of the mean from 1000 trajectories.
The parameters used are $\gamma_c/\gamma_a$=0.5,  $\gamma_m/\gamma_a$=2,  $\eta/\gamma_a$=5,  $V_0/\gamma_a$=20, $V_1/\gamma_a$=40, $\omega_r/\gamma_a$=1, $\Delta_c/\gamma_a$=-1,  $\Delta_a/\gamma_a$=-2, $g_{ac}/\gamma_a$=2, and $g_{mc}/\gamma_a$=2.5.}\label{FIG_QT}
\end{figure}

We demonstrate the computation of $\langle n_c(t)\rangle$ with the QT method (using 1000 trajectories) in Fig.~\ref{FIG_QT}(a) and (b), where the initial states are taken as $|100\rangle$ and $|110\rangle$, respectively. 
It can be seen that the calculations from the QT method are close to that from the quantum master equation (solid black curves), as expected. 
The ratio of the mean excitation in panel (a) to panel (b) is simply the correlation function $G^{(2)}(\tau)$ (will be properly introduced later, see Eq.~(\ref{EQ_g22})), where $t$ is taken to be zero.
For this example, $G^{(2)}(\tau)$ is plotted in panel (c), where it oscillates around unity.

\subsection{The correlation functions}
The correlation functions are normally used to describe coherence properties of electromagnetic fields.
Here we shall compute these quantities for the cavity field mode of the atom-optomechanics system.
The first and second order correlation functions are defined, respectively, as 
\begin{eqnarray}
G^{(1)}(\tau)&=&\frac{\langle \hat a^\dagger(t+\tau) \hat a(t)\rangle}{\sqrt{ \langle {n}_c(t)\rangle \langle {n}_c(t+\tau)\rangle}}, \label{EQ_g1}\\
G^{(2)}(\tau)&=&\frac{\langle \hat a^\dagger(t)\hat a^\dagger(t+\tau)\hat a(t+\tau)\hat a(t)\rangle}{\langle{n}_c(t)\rangle \langle{n}_c(t+\tau)\rangle}, \label{EQ_g2}
\end{eqnarray}
where $\langle {n}_c(t)\rangle =\text{Tr}\left[\hat a^\dagger \hat a \rho(t) \right]$ and $\langle{n}_c(t+\tau)\rangle=\text{Tr}\left[\hat a^\dagger \hat a \rho(t+\tau)\right]$. 

To calculate the numerator of the first order correlation function $G^{(1)}(\tau)$ in Eq.~(\ref{EQ_g1}), the initial density matrix $\rho(0)$ is evolved to $\rho(t)$ with the quantum-classical coupled dynamics.
The subsequent evolution requires helper states, defined as
\begin{eqnarray} 
\tilde{\rho}_1(t)&=&(\mathbbm{1}+\hat a)\rho(t)(\mathbbm{1}+\hat a^\dagger), \nonumber \\
\tilde{\rho}_2(t)&=&(\mathbbm{1}-\hat a)\rho(t)(\mathbbm{1}-\hat a^\dagger), \nonumber \\
\tilde{\rho}_3(t)&=&(\mathbbm{1}+i\hat a)\rho(t)(\mathbbm{1}-i\hat a^\dagger), \nonumber \\
\tilde{\rho}_4(t)&=&(\mathbbm{1}-i\hat a)\rho(t)(\mathbbm{1}+i\hat a^\dagger).
\end{eqnarray} 
Note that this way, we have $(\tilde{\rho}_1(t)-\tilde{\rho}_2(t)-i\tilde{\rho}_3(t)+i\tilde{\rho}_4(t))/4=\hat a\rho(t)\equiv \hat A(t)$.
The normalised helper states (${\rho}_j(t)=\tilde{\rho}_j(t)/\text{Tr}\left[ \tilde{\rho}_j(t)\right]$) are physical density matrices, which are then evolved from $t$ to $t+\tau$.
With this method, one obtains
\begin{eqnarray}
\hat A(t+\tau)&=&\frac{1}{4}[\tilde{\rho}_1(t+\tau)-\tilde{\rho}_2(t+\tau) \nonumber \\
&&-i\tilde{\rho}_3(t+\tau)+i\tilde{\rho}_4(t+\tau)].
\end{eqnarray}
Finally, the first order correlation function is given by 
\begin{equation}
G^{(1)}(\tau)=\text{Tr}\left[\hat a^\dagger \hat A(t+\tau)\right]/\sqrt{\langle n_c(t)\rangle \langle n_c(t+\tau)\rangle}.
\end{equation}

The second order correlation function is computed in a similar way.
After the first evolution, leading to $\rho(t)$, one considers a photon-subtracted state ${\rho_p}(t)= \hat a \rho(t)\hat a^\dagger/\langle{n}_c(t)\rangle$.
This physical state is then evolved from $t$ to $\tau$, giving ${\rho_p}(t+\tau)$. 
The correlation function in Eq.~(\ref{EQ_g2}) is then evaluated as
\begin{equation}\label{EQ_g22}
G^{(2)}(\tau)=\text{Tr}\left[\hat a^\dagger \hat a {\rho_p}(t+\tau)\right]/\langle n_c(t+\tau) \rangle.
\end{equation}
Recall that the denominator in Eq.~(\ref{EQ_g22}) is simply $\text{Tr}\left[\hat a^\dagger \hat a {\rho}(t+\tau)\right]$.
Therefore, it is expected that in the regular phase, given large $\tau$, the state $\rho_p(t+\tau)=\rho(t+\tau)$ is the steady state solution, making $G^{(2)}(\tau)=1$.
This is not the case for the time crystal and chaotic phases, as the mean excitation still oscillates for large $\tau$. 
In this case, the $G^{(2)}(\tau)$ will also oscillate and cross unity during its evolution.

\begin{figure}[h!]
\centering
\includegraphics[width=0.5\textwidth]{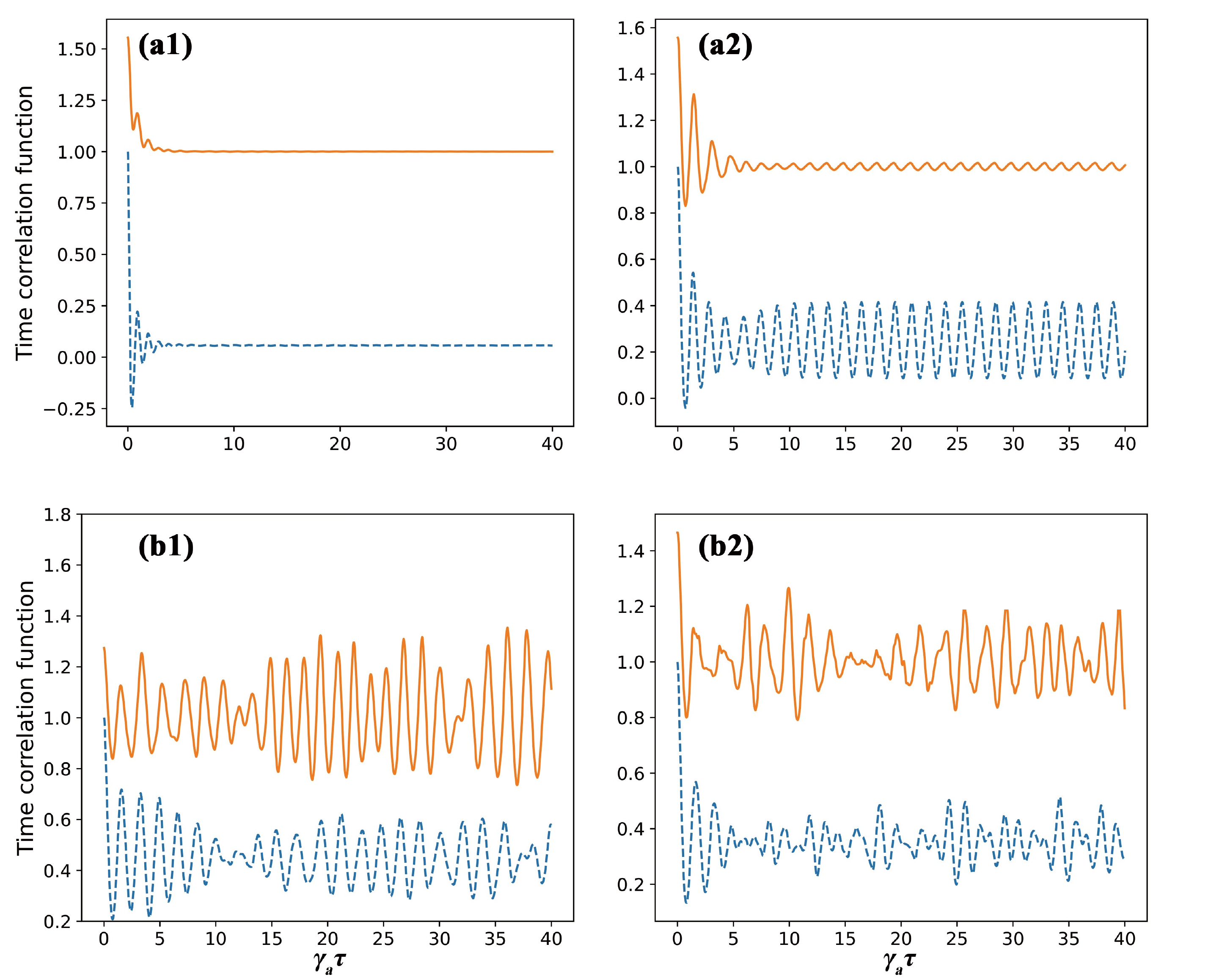}\\
\caption{The first and second order correlation functions of the cavity mode, indicated by the dashed blue and solid orange curves, respectively.
The parameters are summarised as follows $\gamma_c/\gamma_a$=0.5,  $\gamma_m/\gamma_a$=2,  $\eta/\gamma_a$=5,  $V_0/\gamma_a$=20, $V_1/\gamma_a$=40, $\omega_r/\gamma_a$=1, $\Delta_c/\gamma_a$=-1,  $\Delta_a/\gamma_a$=-2, and for panel (a1): $g_{ac}/\gamma_a$=4, $g_{mc}/\gamma_a$=2; (a2): $g_{ac}/\gamma_a$=2, $g_{mc}/\gamma_a$=2.5; (b1): $g_{ac}/\gamma_a$=2, $g_{mc}/\gamma_a$=2; and (b2): $g_{ac}/\gamma_a$=4, $g_{mc}/\gamma_a$=2.}\label{cf}
\end{figure}

As exemplary cases, we present the first (dashed blue curves) and second (solid orange curves) order correlation functions in Fig.~\ref{cf}. 
It can be seen that $G^{(2)}(\tau)\rightarrow 1$ in the regular phase, Fig.~\ref{cf}(a1), while it is oscillating around one in the time crystal phase, as shown in Fig.~\ref{cf}(a2).
This oscillation is also observed in the chaotic phase, but it is random, see Figs.~\ref{cf}(b1)-(b2).

\subsection{The phase diagram for the regular and chaotic phase transition tests}

\begin{figure}[h!]
\centering
\includegraphics[width=0.45\textwidth]{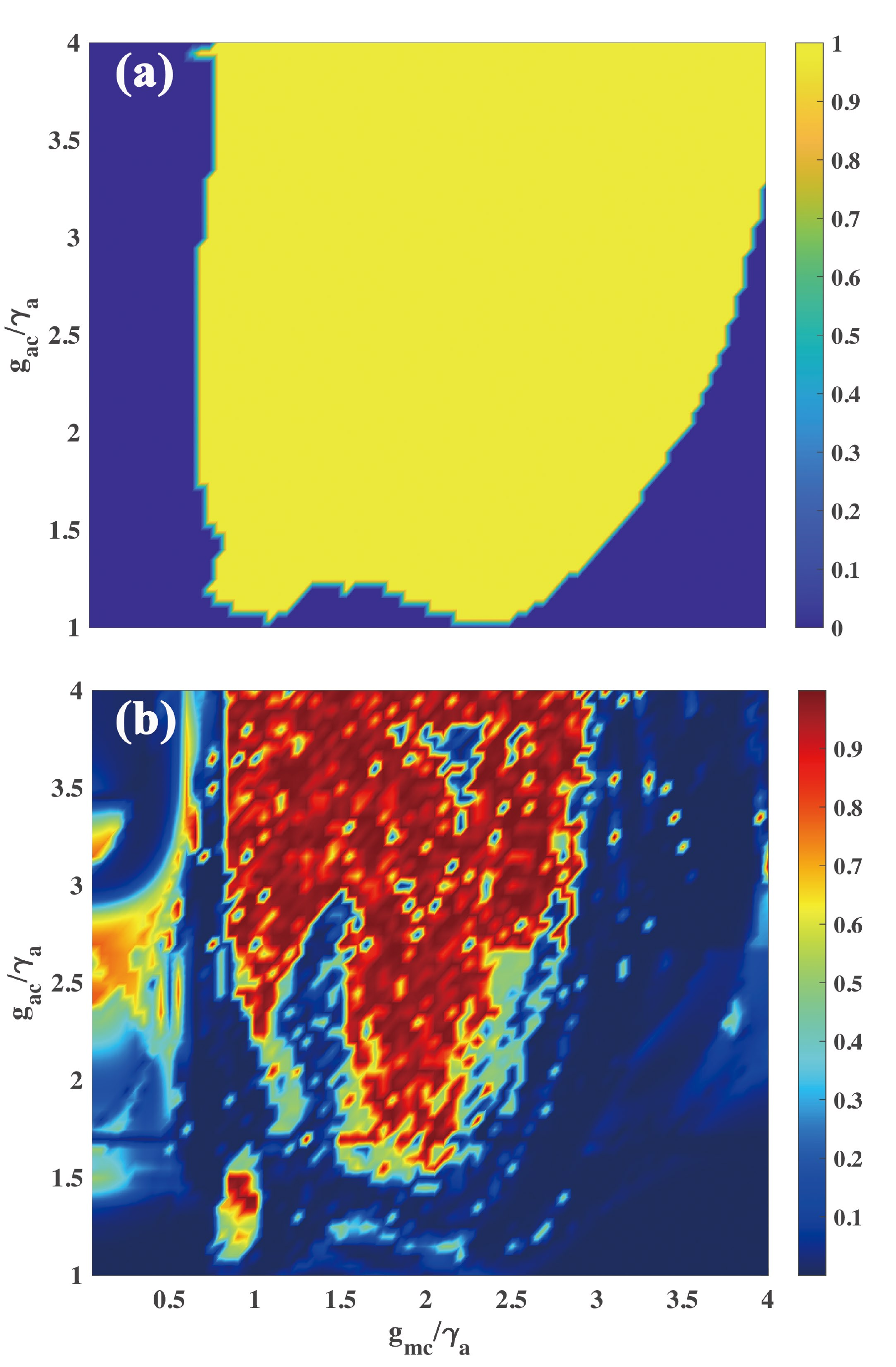}
\caption{The regular phase transition test (a) and the correlation methods for the chaos test (b).  
The parameters used are $\gamma_c/\gamma_a$=0.5, $\gamma_m/\gamma_a$=2, $\eta/\gamma_a$=5, $V_0/\gamma_a$=20, $V_1/\gamma_a$=40, $\omega_r/\gamma_a$=1, $\Delta_c/\gamma_a$=-1, and $\Delta_a/\gamma_a$=-2. }\label{phad2}
\end{figure}

The phase diagram in the main text (Fig. \ref{phad}) is determined by two tests. 
The first one is the regular transition test where it recognizes the cavity mode's excitation converging to a certain value after a long evolution time.
The second one is the chaos test, characterised by Eq.~(\ref{corm}), where $K_c$ will be close to $1$ when the system is in a chaotic phase.

The regular phase transition is shown in Fig.~\ref{phad2} (a) with the blue region indicating the regular phase and the yellow one representing other phases.
When the field-membrane coupling strength $g_{mc}$ is small, there is only regular phase regardless of the field-atom coupling strength $g_{ac}$.
Along with the increase of $g_{mc}$, the system can be in a chaotic or time crystal phase.  However, if $g_{mc}$ is too large, the influence of the atom in the time dependent Hamiltonian can be ignored and the system returns to the regular phase.  

The value of $K_c$ is shown in Fig. ~\ref{phad2} (b) and the chaotic phase, indicated by the redder region, can only exist in the yellow region of Fig.~\ref{phad2} (a).
The chaotic phase appears in the region where $g_{ac}/\gamma_a>1$ and $g_{mc}/\gamma_a\gtrsim1$, i.e., the membrane and the atoms both influence the the dynamical behavior of the whole system.   

The sum of the values from Fig.~\ref{phad2} (a) and Fig.~\ref{phad2} (b) is plotted in Fig. \ref{phad} of the main text.
The  region I (blue color) indicates the regular phase and the region III the chaotic phase (red color). 
The time crystal phase is then inferred from the remaining region, labelled II (green color).

\bibliography{AOM.bib}

\end{document}